# The effect of germanium sublayer on the native corrosion of ultrathin copper films

Vladimir A. Vdovin, Ivan I. Pyataikin[*]

Kotelnikov Institute of Radio Engineering and Electronics of Russian Academy of Sciences, Mokhovaya 11, building 7, Moscow 125009, Russia

Abstract

To examine the process of native corrosion of ultrathin (about 10 nm) copper films deposited on quartz glass substrates ($SiO_2$) with and without a germanium sublayer, the time dependences of the microwave reflection coefficient $R$ and direct current electrical resistivity $\rho$ of such samples exposed to air at room temperature have been studied. Under these conditions, the thickness of the oxide layer $d$ on $Cu/SiO_2$ films was found to increase over time $t$ according to a parabolic law, which is in contradiction with the predictions of existing theories of copper oxidation. A model is proposed that explains this behavior of $d(t)$ by the diffusion of atomic oxygen along the boundaries of oxide grains towards the copper film with its subsequent oxidation. The $R$ and $\rho$ of $Cu/Ge/SiO_2$ films were found to degrade much more slowly than similar characteristics of $Cu/SiO_2$ films of the same thickness. The high corrosion resistance of $Cu/Ge/SiO_2$ films is explained by the peculiarities of Ge redistribution during the growth of the copper film on a germanium sublayer. The long-term retention by $Cu/Ge/SiO_2$ films of their characteristics allows them to be recommended as a cheap replacement for gold coating in electromagnetic interference protection devices.

[*] Corresponding author.
E-mail address: iip@cplire.ru (I. Pyataikin).

# 1. Introduction

It is well known that corrosion processes have a significant impact on the functioning of large-scale metal structures: pipelines, bridges, and the like. For metal films with a thickness of about 10 nm, oxidation and associated corrosion are critical factors that determine and fundamentally limit the lifetime of many useful things containing these films. Electromagnetic interference (EMI) protection devices based on ultrathin metal layers are one of the important examples of this kind of things.

The development of such protective devices began quite a long time ago, immediately after the widespread use of radars [1]. The use of thin metal films for protection purposes is based on the fact that the coefficient of microwave reflection by them becomes close to unity starting from their thickness of about 10 nm [2]. For example, continuous copper films with a thickness of 7.5 nm have a microwave reflectivity greater than 90%, and this value is quite sufficient for effective protection against microwave radiation.

Sputtering films of such small thickness onto clear plastic or glass substrates reduces the transparency of such structures in the visible rather slightly, which allows the successful use of thin metal films for the purpose of protecting human eyes and sensitive optical-electronic equipment from powerful microwave radiation. Throughout this article, we mean precisely this property when we talk about the protective properties of metal films.

For quite a long time, gold remained almost the only material for the production of transparent protective coatings, since it is resistant to oxidation by atmospheric oxygen under normal conditions. Gold is widely used in the electronics industry; its deposition on various types of amorphous, crystalline and plastic surfaces is well mastered. The only drawback of gold films is, naturally, their cost.

As long as the scope of EMI protection devices was rather narrow and limited mainly to military applications, cost issues were of little relevance due to the small number of protective devices. But today the situation has changed dramatically due to the transition to increasingly high-frequency fifth-generation electronic communications technologies and their growing penetration into all areas of everyday life. The explosive growth in the number of 5G base stations and their placement in urban environments with high population density requires orders of magnitude more protective devices than previously available. For this reason, from an economic point of view, it is desirable to use non-precious materials to produce coatings for EMI shielding applications [3-5]. Thus, the search for cheap metals suitable for producing continuous films of nanometer thickness, and at the same time resistant to environmental influences, is a very urgent task.

We were guided by these considerations when we began studying the influence of the atmosphere on the degradation of the protective properties of ultra-thin copper films. For this purpose, we studied the change over time in the microwave reflectivity of copper films deposited on quartz glass substrates and exposed to air at room temperature. Since the reflection of microwaves, and with it the protective properties of the films, are determined by the direct current (dc) conductivity of them, the time evolution of this characteristic was also studied.

Our choice of copper as a material for protective coatings was determined by the fact that copper is an inexpensive material that is extremely widely used in modern microelectronics. Copper films on silicon or glass become continuous starting from a thickness of just over 6 nm [6]. At smaller thicknesses, copper films are island-like, they do not reflect microwaves, and therefore they do not have protective properties.

In a humid air environment, thin copper films of about 10 nm thickness corrode intensively, turning into a mixture of oxides $Cu_2O$ and $CuO$ [7] that conduct electrical current poorly. For this reason, such films quickly lose dc conductivity, and with it their protective properties. On the other hand, at room temperature, copper layers several hundred nanometers thick are quite resistant to oxidation. The thickness of the oxide layer on them grows logarithmically slowly over time and does not exceed 10 nm after four months from the moment of contact with the atmosphere [8].

Copper oxidation has been studied extensively for over 100 years. A comprehensive review of theoretical and experimental works on this issue can be found in Refs. [9 - 13] and in the introductory sections of articles [14 - 16]. An analysis of the literature references cited in these publications shows that virtually all experimental studies devoted to the growth of oxide films on Cu at room temperature examined fairly thick copper layers with a thickness of more than several hundred nanometers. It is generally accepted that the growth of the oxide film on them is due to the diffusion of $Cu^+$ ions through $Cu_2O$ oxide layer towards the air/oxide interface, where the reaction of $Cu^+$ with atomic oxygen occurs. The theories of transport of $Cu^+$ ions through the oxide layer listed in Refs. [9 - 16] lead to time dependences of the thickness $d$ of the oxide film according to the logarithmic $d = B\,ln(At + 1)$ [17] or inverse logarithmic law $d^{-1} = B - A\,ln(t)$ [18], where $A$ and $B$ are some characteristic constants.

On the contrary, in our work we found that at room temperature the growth of oxide on copper films with a thickness of about 10 nm follows a parabolic law, $d^2 \propto t$. The model we propose explains this dependence of $d(t)$ by the diffusion of atomic oxygen along the boundaries of $Cu_2O$ grains in the direction from the air/oxide interface to the surface of the copper film.

As noted above, copper films acquire their protective properties starting from a thickness slightly greater than 6 nm. However, for a number of applications, continuous copper layers of much smaller thickness may be required. One of the methods for obtaining such films is the deposition of copper on a germanium intermediate layer (sublayer). Copper films grown on such a sublayer ("Cu/Ge films" for short) become continuous already at a thickness of 1.5 nm [19].

By studying the time evolution of microwave reflectivity of copper films several tens of nanometers thick, grown on a germanium sublayer 1.5 - 2 nm thick, we found that such films degrade significantly more slowly than Cu films of the same thickness, but grown directly on a glass substrate. In our work, we explain this practically important fact by the peculiarities of Ge redistribution during the growth of copper films on a germanium sublayer.

Despite the fact that the characteristic time of retention of the protective properties by Cu films grown on a germanium sublayer is significantly longer than the similar parameter for conventional copper films, at some point Cu/Ge films also begin to degrade. We found that the process of their degradation is quite unusual, and its rate strongly depends on the ratio of the thicknesses of the copper and germanium layers. If these thicknesses are comparable, the dc conductivity of the Cu/Ge film drops by half in about a month. And in the case where the thickness of the Cu layer is four times greater than the thickness of the Ge sublayer, the same time is extended to a year.

In the article we show that such a degradation mode of the protective properties of Cu/Ge films is associated with the high diffusion activity of germanium and its transport mainly along the intercrystalline boundaries of the copper film. Understanding these features allowed us to formulate recommendations for obtaining Cu/Ge EMI shielding coatings that retain their protective properties for a long time.

## 2. Samples and methods

The deposition of Cu and Ge layers was carried out by thermal evaporation of copper (99.999% pure) and germanium (99.9999%) in a vacuum of $5 \cdot 10^{-6}$ mm Hg onto substrates of quartz glass polished to optical quality, which were at room temperature at the time of deposition and pre-annealed at 250°C. The choice of quartz glass as the substrate material for all the samples studied was determined by its transparency in the microwave frequency range due to the negligible dc conductivity of this type of glass at room temperature.

The surface morphology of deposited Cu and Cu/Ge films was studied by using scanning electron microscopy. X-ray reflectometry was employed to control the thickness of copper and germanium layers.

The study of degradation processes of Cu and Cu/Ge films in our work was performed by using two methods. The first was based on the study of the change in microwave reflection, transmission and absorption coefficients of the films over time; the second relied on examining the evolution of their dc conductivity.

Two types of samples were made for microwave measurements. The samples of the first type had the Cu film deposited directly onto a glass substrate with dimensions of 22,9×9,8×4 mm³. And the second type of samples had a design where the copper film was deposited on a 1.9 nm thick germanium layer immediately after its condensation on a glass substrate of the same dimensions. Measurements of the microwave coefficients of these samples in the frequency range of 8.5 – 12.5 GHz were carried out in a rectangular waveguide with a cross-section of 23×10 mm² according to the scheme described by us earlier in Ref. [20]. The microwave coefficients measured in this way were normalized to their initial value at $t = 0$.

For dc measurements, five samples were made, two of which had a 6.5 nm thick Cu film evaporated directly onto substrate, while the other three had copper films of 2.1, 4.8 and 7.3 nm thickness deposited onto substrates with a 2 nm thick germanium sublayer. In all these cases, glass substrates were employed that were completely similar to those used to make samples for microwave measurements, but that were half as thick (2 mm).

To connect these samples to the measuring circuit, narrow (approximately 3 mm) copper strips with thicknesses of 40 – 50 nm were deposited on the surface of the films being studied near the face ends of the substrates, and a pair of millimeter-sized pieces of indium was attached to each such strip by a simple pressing. The structures obtained in this way allowed conductivity of the films to be measured using the standard four-probe method.

## 3. Results and discussion

### 3.1. Analysis of experimental data

Figure 1 shows the change over time in the normalized microwave reflection coefficient, $R/R_0$, from two types of copper films. The monotonic decrease in the $R/R_0$ of films deposited directly onto a glass substrate is due to the steady decrease in their conductivity. In copper films grown on a germanium sublayer, the degradation of the normalized reflection coefficient, and consequently the film conductivity, is practically unnoticeable over a time interval of about 1000 hours.

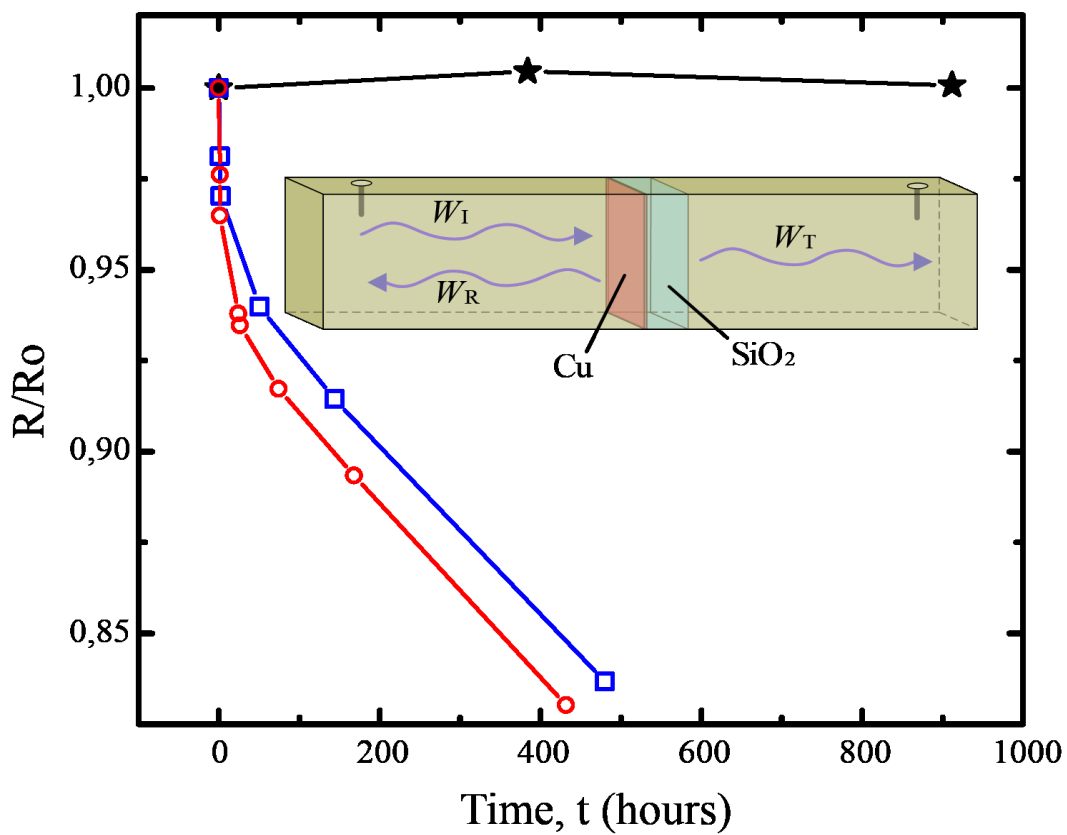

**Fig. 1.** Time evolution of normalized microwave reflection coefficients $R/R_0$ for 7.9 nm thick copper films deposited directly on a quartz glass substrate (squares and circles) and for a 7.3 nm thick Cu film grown on a germanium sublayer with a thickness of 1.9 nm (stars). $R = W_R/W_I$, where $W_R$ and $W_I$ are the powers of microwave radiation reflected by the film and incident on it, respectively. $R_0$ is the initial value of $R$: $R_0 \equiv R(t=0)$. All the coefficients shown in the figure were measured at a frequency of 10 GHz. The magnitude of the errors in determining $R/R_0$ is less than the size of the symbols representing the experimental data. Slightly different degradation rates of identical Cu films are due to different ambient air humidity. Lines connecting experimental points are drawn to facilitate the perception of the data. The inset shows the measurement diagram.

As can be seen from the figure, copper films deposited directly onto glass lose their protective properties much faster than similar films grown on a germanium sublayer. Therefore, Cu/Ge films are much more resistant to environmental influences than simple copper films, and this practically important fact should be studied in detail to establish its causes.

Although microwave measurements allowed us to detect high corrosion resistance of Cu/Ge/SiO$_2$ films, it should be pointed out that the experimental dependences $R(t)/R_0$ shown in Fig. 1 are quite difficult to use to determine the reasons underlying this property, as well as the mechanisms causing rapid degradation of the conductivity of simple copper films. This is due to the fact that the dependence of $R/R_0$ on the film conductivity is extremely complex owing to the influence of reflection of microwaves from the substrate on the reflection coefficient of the entire structure.

For this reason, to clarify the mechanisms of degradation of the films, it is preferable to proceed to the analysis of the time dependences of their dc conductivity $\sigma$ or specific electrical resistance $\rho = 1/\sigma$, which can be measured relatively easily and do not contain contributions from the dielectric substrate. Such dependences are presented in Fig. 2.

As can be seen from Fig. 2 (a), the conductivity of simple copper films decreases monotonically over time and after two days drops by almost a third. This time dependence of conductivity is consistent with the steady decrease in the microwave reflection coefficients of copper films deposited directly on glass substrates, shown in Fig. 1.

On the contrary, as follows from Fig. 2 (b), the conductivity of films grown on a germanium sublayer changes non-monotonically, and during the first 20 days some improvement in the conductivity is even clearly visible compared to the starting value. As can be seen from this figure, after 80 days the change in conductivity is less than 4% of the initial value. All this is also in excellent agreement with the microwave data of Fig. 1.

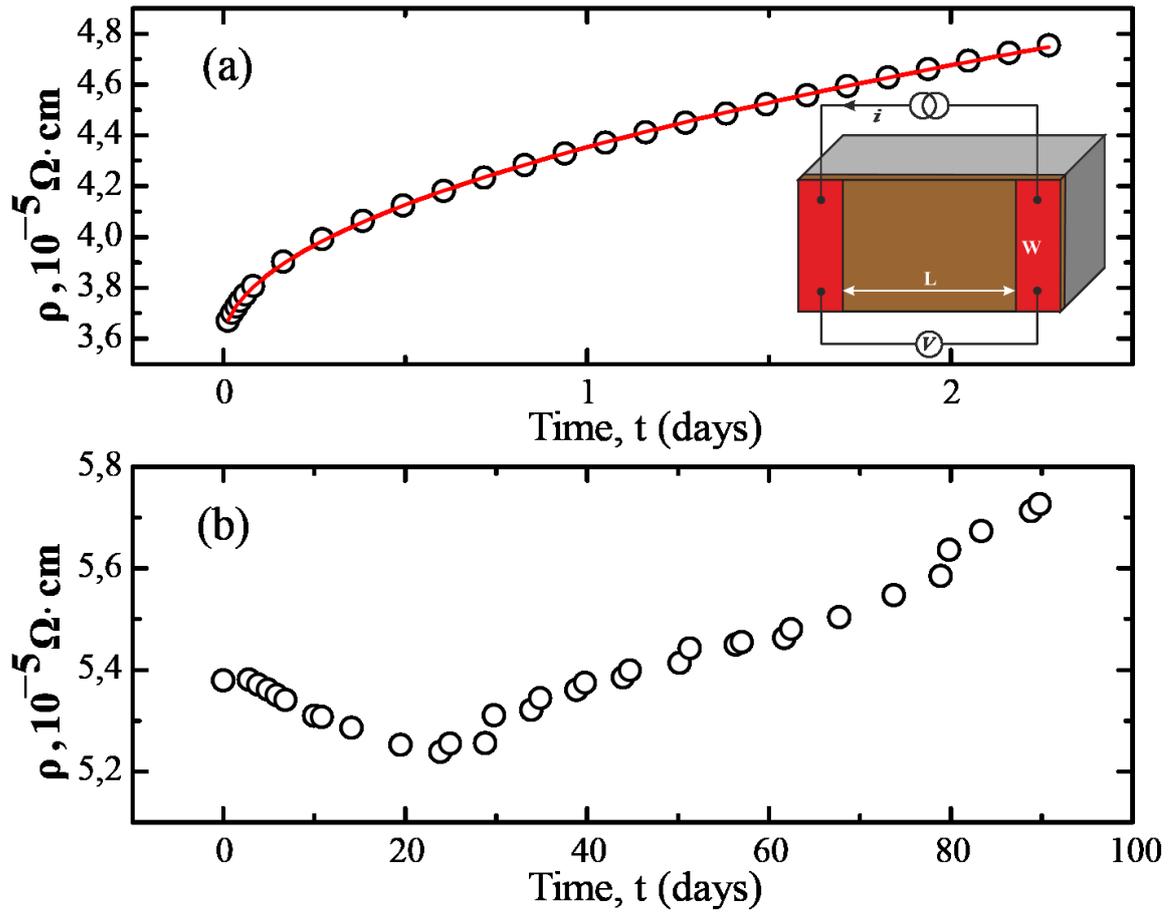

**Fig. 2.** (a) $\rho(t)$ dependence for a 6.5 nm thick copper film deposited directly on a quartz glass substrate. The solid line is the result of fitting the function $a + bt^\beta$ to experimental points. The exponent $\beta$ obtained from this fitting is $0.506 \pm 0.004$. The inset shows the basic diagram of the conductivity measurement. (b) $\rho(t)$ for a 7.3 nm thick copper film grown on a 2 nm thick germanium sublayer.

Thus, based on the data presented in Fig. 1 and 2, it can be stated that the conductivity, and with it the microwave coefficients of Cu/Ge films, degrade much more slowly than similar characteristics of Cu films of the same thickness, but deposited directly on quartz glass substrates. The reasons for the high corrosion resistance of Cu/Ge films are discussed further in Section 3.3.

Figure 3 shows the $\rho(t)/\rho_0$ dependences for three Cu/Ge samples, in which copper films with thicknesses of 2.1, 4.8 and 7.3 nm were deposited on top of a germanium sublayer with a thickness of about 2 nm. As follows from the figure, the region of steady degradation of conductivity of all three samples is preceded by an initial part in which some improvement in the conductivity is noticeable compared to the value at $t = 0$.

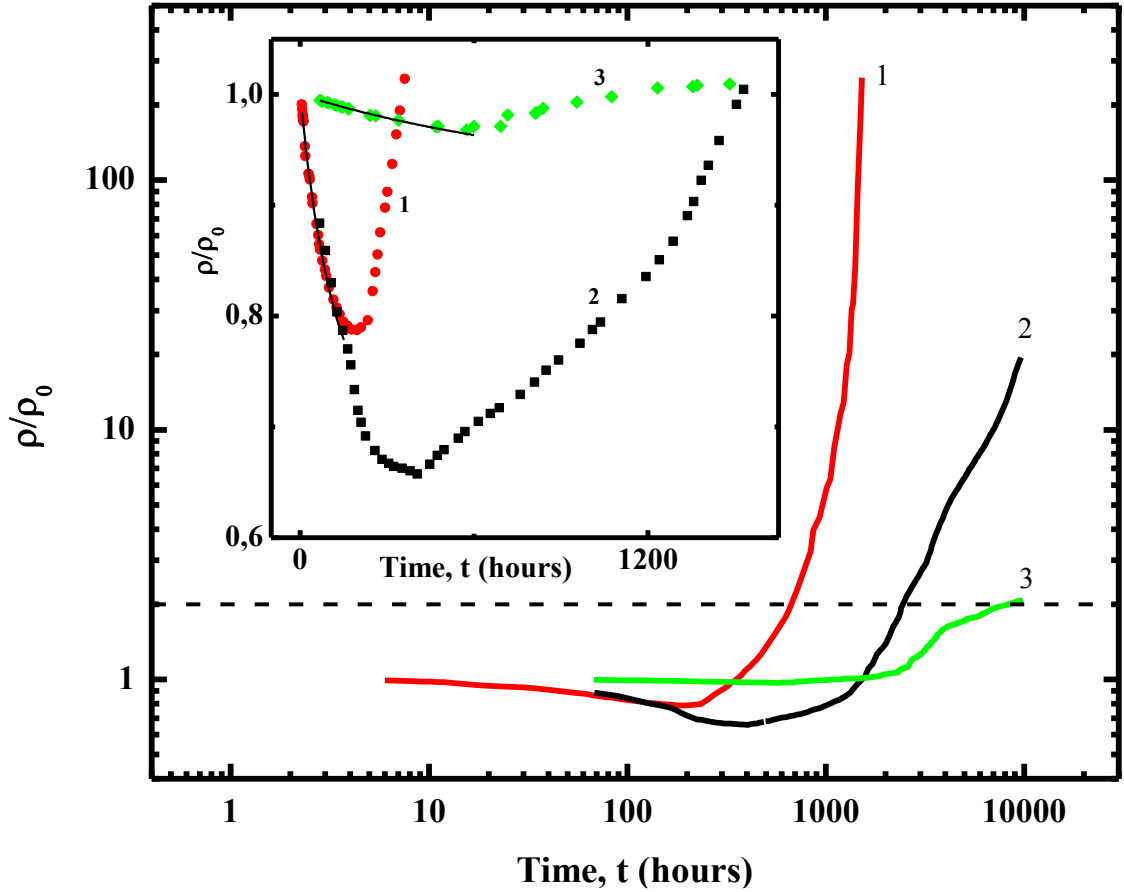

**Fig. 3.** Curves 1–3: $\rho(t)/\rho_0$ dependences for three Cu/Ge samples with copper layer thicknesses of 2.1, 4.8 and 7.3 nm, deposited on a germanium sublayer with a thickness of about 2 nm. $\rho_0 \equiv \rho(t=0)$. The dashed line corresponds to a twofold decrease in the sample conductivity. Inset: A more detailed view of the initial part of all three dependences. The solid lines in the inset are the result of fitting equation (5) to the experimental points. The values of the adjustable parameters $g = G_\infty/G_0$ and $\mathcal{K}_a$ obtained from the fitting are given in Table 1.

The rate of conductivity degradation, as can be seen from the figure, depends on the ratio of the thicknesses of the copper film and the germanium sublayer. If these thicknesses are approximately the same, then the conductivity of the Cu/Ge sample drops by half in about a month (28 days). In the case where the thickness of the copper layer is approximately twice the thickness of the sublayer, the same time increases to three months (98 days). If the thickness of the copper film is approximately four times greater than the thickness of the sublayer, the time for a twofold decrease in the sample conductivity is extended to almost a year (11 months).

As follows from the figure, the sample with the largest ratio of the thicknesses of the copper film and the sublayer demonstrates a peculiarity of the conductivity degradation rate. Namely, starting from a certain point in time, the rate of loss of conductivity by the sample abruptly slows down, so a distinct break is noticeable on the $\rho(t)/\rho_0$ curve. The reasons for such an unusual mode of degradation of the conductivity of Cu/Ge films and its dependence on the ratio of the thicknesses of the copper and germanium layers are clarified further in section 3.3.

As noted in the Introduction, the slow increase in the thickness of the oxide layer and the resulting decrease in the conductivity of copper layers with a thickness of more than a few hundred nanometers follows a logarithmic law. At the same time, it can be seen from Fig. 2 (a) that the degradation of the conductivity of thin (about ten nanometers) copper films occurs at a much faster rate. Indeed, the experimental data presented in this figure are fitted with a high degree of accuracy by an empirical power-law dependence of the form $a + bt^\beta$ with an exponent $\beta = 0{,}506 \pm 0{,}004$.

The closeness of the fitting parameter $\beta$ to ½ allows us to assume that the thickness of the oxide layer on thin copper films changes over time according to the parabolic law $d(t) \propto \sqrt{t}$. In this case, the decrease in the thickness of the metal part of the film $h$ during the growth of the oxide layer is obviously described by the expression $h = h_0 - \sqrt{k_p t}/\gamma$, where $h_0$ is the initial thickness, $k_p$ is the growth constant of the oxide layer, $\gamma$ is the ratio of the increase in the thickness of the oxide layer to the decrease in the thickness of the metal part of the film caused by it. As will be shown in the next section, for copper films $\gamma \approx 1{,}64$.

If we neglect the increase in $\rho$ of the film with a decrease in its thickness $h$ caused by the classical size effect, then for the electrical resistance of the film per square $R^\square$ we can write the expression

$$R^\square \equiv \rho/h \approx \rho\left(1 + \sqrt{k_p t}/(\gamma h_0)\right)/h_0 \equiv R_0^\square\left(1 + \sqrt{Kt}\right), \qquad (1)$$

which is valid as long as $\Delta h = \sqrt{k_p t}/\gamma \ll h_0$.

The significance of this equation is twofold. Firstly, it shows that the growth of the oxide film thickness according to the parabolic law does indeed lead to the experimentally observed variation of $\rho(t)$. Secondly, the obtained expression (1) depends on only two adjustable parameters $R_0^\square$ and $K$, and therefore can be used to obtain a reliable estimate of the oxide film growth constant $k_p$ by fitting this expression to the experimental dependences $R^\square(t)$.

In Fig. 4, the solid line shows the result of such a fitting of the initial part of the experimental dependence $R^\square(t)$. A fairly good match between the experimental points and the curve calculated using formula (1) allows us to obtain an estimate of $k_p$ with an accuracy of about 1%: $k_p = (\gamma h_0)^2 K = (6{,}85 \pm 0{,}05) \cdot 10^{-19}$ cm²/s.

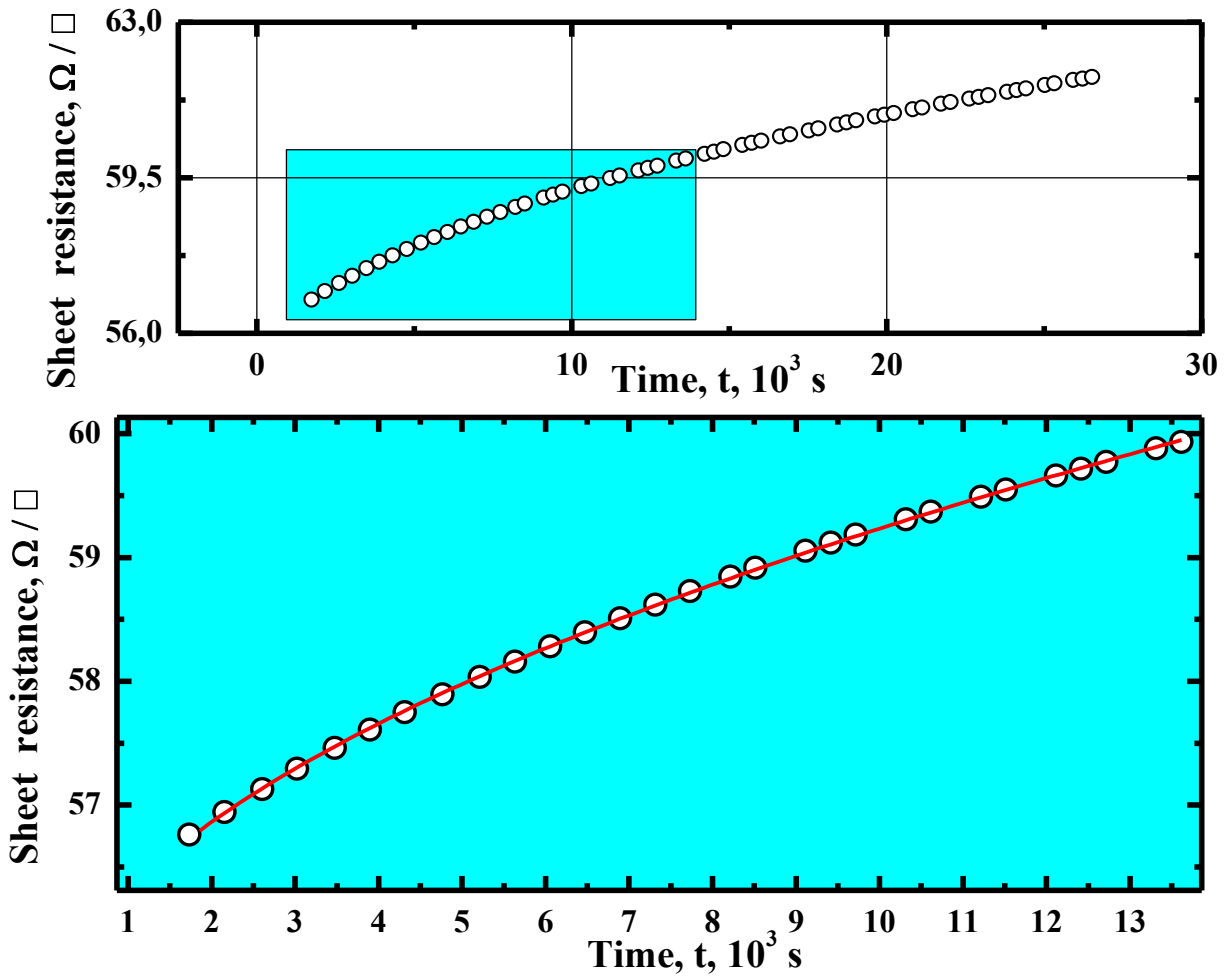

**Fig. 4.** The time dependence of the electrical resistance of Cu film per square $R^\square = R\, L/w$. Here $R$ is the film resistance measured using the four-contact method, $L$ and $w$ are the length and width of the film, respectively (see inset to Fig. 2a). The film thickness $h_0$ is 6.5 nm. The initial part of the experimental dependence $R^\square(t)$ is shown at the bottom on an enlarged scale. The solid line is the result of fitting equation (1) to the experimental points with adjustable parameters $R_0^\square = 54{,}96 \pm 0{,}01$ Ω and $K = (6{,}03 \pm 0{,}04) \cdot 10^{-7}$ s⁻¹.

Thus, based on the data presented in Fig. 2 (a) and 4, it can be stated that at room temperature the thickness of the oxide layer $d$ on the ultrathin Cu films deposited directly on quartz glass, increases with time according to a parabolic law. This fact contradicts the established consensus, according to which the parabolic growth mode of the oxide film should be observed at temperatures greater than 180 °C, since at lower temperatures the diffusion coefficient of oxygen and Cu ions in the oxide layer is small, and at room temperature it is negligibly small. We propose a resolution to this contradiction in the next section.

*3.2. Degradation of conductivity of Cu thin films*

As mentioned above, the interaction of copper with oxygen leads to the formation of a $Cu_2O$ oxide film, a new phase at the air/copper interface. The growth process of this oxide can be divided into two stages: the appearance and growth of individual $Cu_2O$ islands and the subsequent stage of growth of a continuous layer of this oxide formed after the merging of the islands. The key process at the first stage is the adsorption of the $O_2$ molecule on the Cu surface, followed by dissociation into a pair of oxygen atoms that then react with copper to form the $Cu_2O$ molecule. The most active dissociation processes occur in the vicinity of growth steps and other structural imperfections on the surface of the copper layer, where the values of the electric field arising owing to the appearance of a charge at the interphase boundary are especially high.

It should be pointed out that the stage of $Cu_2O$ island growth is the moment when the process of copper corrosion can be influenced with maximum efficiency. Since further reactions occur on the interface between two solid phases, by eliminating or making the formation of one of them (in the case under consideration, $Cu_2O$) as difficult as possible, it is possible to eliminate or greatly weaken the corrosion process as a whole. We will return to this possibility in Section 3.3 when discussing the anomalously low rate of conductivity degradation of copper thin films grown on a germanium sublayer.

The rate of $Cu_2O$ growth at the island stage is directly proportional to the number $j_{O_2}$ of $O_2$ molecules falling on a unit surface of the copper layer per unit time. Considering air to be an ideal gas, in which the velocities of molecules are subject to the Maxwell distribution, the flow $j_{O_2}$ can be estimated by the relation $j_{O_2} = p/\sqrt{2\pi m k_B T}$, following from elementary kinetic theory. (Here $m$, $k_B$ and $T$ are the mass of the oxygen molecule, Boltzmann's constant, and the absolute temperature, respectively). Substituting the partial pressure of oxygen in the air ($0{,}21 p_0$) instead of $p$ into this expression, we obtain that under normal conditions ($p_0 = 101325$ Pa, $T_0 = 295$ K) $5.8 \cdot 10^{22}$ $O_2$ molecules fall per second on each square centimeter of the copper layer surface.

In view of that the density of atoms on the surface of the copper layer $n_{Cu}$ is of the order of $1/a_{Cu}^2 \approx 7.7 \cdot 10^{14}$ cm$^{-2}$ (here $a_{Cu} = 3.61496$ Å [21] is the unit cell parameter of Cu), we come to the conclusion that even taking into account the low value of the sticking coefficient of molecular oxygen to copper ($S_{O_2} \cong 10^{-2} \div 10^{-3}$, according to the data of work [22]), a continuous $Cu_2O$ layer is formed under normal conditions in a time of the order of $n_{Cu}/(S_{O_2} j_{O_2}) \approx 1.3 \cdot 10^{-5} \div 1.3 \cdot 10^{-6}$ s, that is, almost instantly.

Thus, the island stage of $Cu_2O$ growth is extremely short. Just a few microseconds after contact with the atmosphere, a continuous oxide film forms on the surface of the copper layer, which not only hinders the access of the reacting substances to each other, but also fundamentally changes the kinetics of the corrosion process. For example, oxygen dissociation now begins to occur on the surface of the oxide film, rather than on the surface of the copper layer.

Depending on the degree of structural perfection of the metal layer, two mechanisms for further growth of the thickness of the oxide film on it are possible: the classical diffusion mechanism and electron-tunneling mechanism by Mott. The latter mechanism essentially relies on the idea of stimulating the transport of $Cu^+$ ions by an electric field existing inside the $Cu_2O$ layer and arising due to the tunneling of electrons from copper to oxygen atoms formed on the surface of the oxide layer. Since the tunneling probability and the degree of influence of the electric field on the migration of copper ions decrease exponentially with increasing Cu$_2$O layer thickness, the Mott mechanism leads to a rapidly decreasing growth rate of the oxide film and the logarithmic dependences $d(t)$ indicated in the Introduction.

The driving force of the diffusion mechanism of oxide layer growth is the gradient of the concentration of atomic oxygen as well as oxygen and copper defects migrating through the thickness of copper oxide Cu$_2$O. To estimate the rate of growth of the Cu$_2$O film thickness in accordance with this mechanism, it is fundamentally important to know the values of the strongly temperature-dependent diffusion coefficients of atomic oxygen and Cu$^+$ ions travelling through the oxide layer.

To describe the temperature dependence of the diffusion coefficient of impurities or defects through a solid, it is customary to use a two-parameter semi-empirical formula:

$$D(T) = D_0 \exp(-T_a/T), \qquad (2)$$

in which the parameters $D_0$ and $T_a$ are determined from experiment. For example, fitting equation (2) to the experimental points in Fig. 3 from work [23] yields the values of the parameters $D_{00} \approx 1.1 \cdot 10^{-2}$ cm$^2$/s and $T_{a0} \approx 20565 \pm 1845$ K (1.77 ± 0.16 eV), characterizing the diffusion of oxygen atoms as interstitial impurities in crystalline perfect Cu$_2$O layers. The temperature dependence of the diffusion coefficient of copper vacancies (Cu atoms and Cu$^+$ ions) in Cu$_2$O is also described by formula (2) with parameters $D_{01} \approx (9.1 \pm 1.9) \cdot 10^{-4}$ cm$^2$/s and $T_{a1} \approx 14336 \pm 237$ K (1.24 ± 0.02 eV) [24].

Calculation of the diffusion coefficient according to formula (2) using the above values of the coefficients $D_{00}$ and $T_{a0}$ gives at room temperature an extremely small value of $D(295\ K) \approx 6 \cdot 10^{-33}$ cm$^2$/s. Indeed, with such a diffusion coefficient, it would take about a third of the lifetime of the solar system for an oxygen atom to diffuse a distance of the order of the lattice constant of copper oxide ($a_{Cu_2O} = 4.2696$ Å, [21]): $t = a_{Cu_2O}^2/(6D) \cong 1.6 \cdot 10^9$ years.

Calculation under similar conditions of the diffusion coefficient of copper vacancies gives $7 \cdot 10^{-25}$ cm$^2$/s. Although this value is 8 orders of magnitude greater than the diffusion coefficient of oxygen, it still remains small, since it would take a copper ion about 13.5 years to cover a distance of the $a_{Cu_2O}$.

These estimates show that at room temperature, the growth of the oxide film on crystalline perfect copper layers can only proceed by the Mott mechanism, leading to a logarithmically slow growth of the oxide thickness, which is observed experimentally [8, 25].

The situation is completely different during the oxidation of the structurally imperfect thin films that we studied, obtained under conditions of strong supercooling of a metal condensate during its deposition on a substrate being at room temperature. As shown in the previous section, such films under normal conditions demonstrate a diffusion mechanism of growth of the oxide layer thickness, $d(t) \propto \sqrt{t}$. This mechanism is due to the structural features of these films, caused by strong supercooling of the condensate during deposition.

Indeed, it is known that films obtained under such conditions consist of randomly oriented small crystallites (grains), the large misorientation angles between which lead to acceleration of diffusion along the large-angle intergrain boundaries. Therefore, these films are diffusion-active, and this activity is due to the presence of point defects in crystallites as well as pores and a large number of intergrain boundaries [26]. For this reason, the oxide layer growing on the surface of such films is also highly structurally imperfect and saturated with a large number of pores [27], forming a developed system of channels through which the transport of atomic oxygen and copper is facilitated. In order to find out which of these two elements' diffusion makes the main contribution to the growth of the oxide film, one should refer to the literature.

In Ref. [28], the diffusion of atomic oxygen along grain boundaries in copper oxide layers of varying structural perfection was studied (see Fig. 25 of the referred work). From this figure, the constants $D_0$ and $T_a$, characterizing the diffusion transfer of oxygen in $Cu_2O$, can be extracted. Thus, in thick (more than 100 nm) layers with a high degree of crystalline perfection, these constants are equal to $D_{02} \approx 1.6 \cdot 10^{-4}$ cm²/s and $T_{a2} \approx 13483 \pm 168$ K (1.16 ± 0.01 eV). For thinner and less perfect layers, similar constants are $D_{03} \approx 6.1 \cdot 10^{-8}$ cm²/s and $T_{a3} \approx 6203 \pm 282$ K (0.53 ± 0.02 eV).

In Ref. [29] it was established that in the temperature range of 180 - 260 ˚C the time dependence of the thickness $d$ of the oxide layer on copper films deposited on glass substrates in a technical vacuum of $10^{-3}$ Pa follows a parabolic law. From Fig. 5 of work [29] the growth constants $k_p$ can be obtained for five temperatures from the above-mentioned range. Fitting the function $k_p^Z(T) = k_0 \exp(-T_a^Z/T)$ to these experimental data yields $k_0 \approx 1.0 \cdot 10^{-8}$ cm²/s and $T_a^Z \approx 6704 \pm 441$ K (0.58 ± 0.04 eV). The agreement, within the measurement accuracy limits, of the Arrhenius temperature $T_a^Z$ obtained in this case with the similar parameter $T_{a3}$ from article [28] allows us to conclude that the growth of the oxide layer on structurally imperfect copper films obtained by thermal evaporation in Ref. [29] is due to the diffusion of atomic oxygen along the boundaries of $Cu_2O$ grains.

Calculation of $k_p^Z$ at room temperature gives $k_p^Z(295 \text{ K}) = 1.4 \cdot 10^{-18}$ cm²/s, which is approximately twice the experimental value $k_p = 6.9 \cdot 10^{-19}$ cm²/s, that we obtained in section 3.1. However, it can be considered that these values are in complete agreement with each other, taking into account the rather large error in determining the pre-exponential factor $k_0$, caused by the scatter of data in Fig. 5 of work [29]. The closeness of the extrapolated value of $k_p^Z(295 \text{ K})$ to the experimental value of $k_p$ allows us to conclude that, even at room temperature, the growth of oxide on the copper films we studied is due to the diffusion of atomic oxygen along the boundaries of $Cu_2O$ grains.

Summarizing all of the above, we can propose the following model of oxide layer growth on structurally imperfect copper films.

From the moment a continuous layer of $Cu_2O$ oxide is formed, dissociation of molecular oxygen from the atmosphere begins on its surface. The resulting atomic oxygen diffuses along the boundaries of $Cu_2O$ grains from the outer surface of the oxide layer to the copper surface, where it reacts with Cu to form new portions of $Cu_2O$.

Let $r$ denote the ratio of the number of oxygen atoms to the number of $Cu_2O$ molecules in the oxide layer. We assume that the oxide film is completely homogeneous, therefore the process of diffusion of atomic oxygen through it can be considered one-dimensional. Let us define the coordinate axis X across the oxide film, directing it into the oxide, and select the origin on the air/oxide interface. Then, for the diffusion flow $j$ along a given axis, in accordance with Fick's law, we obtain: $j = -D\, dn/dx$, where $n = n(x)$ is the concentration of atomic oxygen, and $D$ is its diffusion coefficient along the boundaries of $Cu_2O$ grains. Since oxygen does not accumulate in the thickness of the oxide layer during the diffusion process, the flow $j$ does not depend on $x$. In this case, integrating the Fick equation given above, we obtain that the concentration $n$ is a linear function of the coordinate $x$: $n(x) = n_1 x + n_0$, where $n_0$ is the concentration of atomic oxygen near the air/oxide interface ($x = 0$). On the other surface of the oxide film ($x = d$), where there is active consumption of oxygen due to its reaction with copper, the oxygen concentration is low, and we consider it equal to 0. From here we immediately find that $n_1 = -n_0/d$. Thus, $n(x) = (d - x)\, n_0/d$. Substituting this expression into Fick's law, we obtain that $j = D\, n_0/d$.

The presence of the diffusion flow $j$ inside the oxide layer means that during time $\Delta t$, on a section of the Cu film surface with an area $S$, $\Delta N_{Cu_2O} = jS\Delta t$ new

Cu₂O molecules arise due to interaction of copper with oxygen. As a result, the thickness of the oxide layer increases by the amount $\Delta d = \alpha \Delta N_{Cu_2O}$, where $\alpha$ is a constant that can be expressed through the material parameters of Cu and Cu₂O. In view of the all relationships written above, the expression for the increase in the thickness of the oxide layer takes the form:

$$\Delta d = \alpha D S \Delta t \, n_0 / d. \quad (3)$$

The concentration $n_0$ can be found from the condition $S \int_0^d n(x) dx = N_o$, where $N_o$ is the total number of oxygen atoms in the oxide layer, which, in turn, can be expressed through its thickness: $N_o = r N_{Cu_2O} = rd/\alpha$, where $N_{Cu_2O}$ is the total number of Cu₂O molecules in the layer. Calculating the elementary integral gives $n_0 = 2r/(\alpha S)$. Substituting this relation into eq. (3) and replacing the increments of both $d$ and $t$ in it with their differentials, we obtain the simplest differential equation, whose integration yields a parabolic law for $d(t)$:

$$d(t)^2 = d_i^2 + 4rDt. \quad (4)$$

Here $d_i$ is the minimum thickness of the continuous oxide layer, on reaching which the above assumptions regarding the nature of the diffusion process of atomic oxygen in the oxide film become valid. The value of $d_i$ is of the order of the thickness of the Cu₂O monolayer. Therefore, in all further estimations we will assume that $d_i \cong a_{Cu_2O} \approx 4.3$ Å.

As follows from eq. (4), the growth constant $k_p = 4rD$ is determined not only by the diffusion coefficient of atomic oxygen $D$, but also depends on the parameter $r$, which characterizes the saturation of the oxide layer with this active form of oxygen. The value of this parameter depends on the temperature and can be determined from the relation $k_p = 4rD$ if the pre-exponential factors in the Arrhenius laws for the diffusion coefficient and growth constant are known.

For example, using the pre-exponential factor $k_p'^0$ of the thermogravimetric growth constant of the oxide layer from article [30] and the value of $D_{00}$ obtained from the data by Moore et al. [23], we find that $r = k'^0_p/(4D_{00}\rho^2_{Cu_2O}) \approx 0.66$ in the temperature range of 800 – 1050 ˚C. And using the above values of $D_{03}$ and $k_0$, extracted from the data of articles [28] and [29], respectively, allows us to estimate the $r$ at lower temperatures (180 - 260 ˚C): $r = k_0/(4D_{03}) \approx 0.014$.

From this it is clear that, firstly, the $r$ increases with increasing temperature, and secondly, that the solution of atomic oxygen in Cu₂O cannot be considered ideal even at moderate temperatures, which are discussed in article [29], not to mention the high temperatures discussed in works [23] and [30, 31]. The latter conclusion seems particularly important, since the non-ideality of the solution means that the ideal law of mass action, one of the cornerstones of chemical thermodynamics, as well as Wagner's theory [32], is inapplicable.

In section 3.1, based on our experimental data, we estimated the constant of the oxide growth on thin copper films obtained under conditions of strong supercooling of deposited metal during its condensation on a substrate that was at room temperature at the time of film deposition: $k_p = (6{,}85 \pm 0{,}05) \cdot 10^{-19}$ cm²/s. With this constant, it is easy to calculate the thickness of the oxide layer appearing in an hour ($\tau = 3600$ s) on the surface of a copper film obtained using a similar technology and exposed to air at room temperature: $\sqrt{d_i^2 + k_p\tau} \approx 7$ Å, which is a completely reasonable value.

It should be pointed out that the above parameter $k_p$ is also very useful for obtaining practically important estimates concerning the corrosion process of structurally imperfect copper films at room temperature. For example, it can be used to calculate the characteristic time $\tau_0$ of complete degradation of a copper film of thickness $h$, which allows one to evaluate the service life of devices that include such Cu layers.

To calculate $\tau_0$, it is necessary, first of all, to determine the thickness of the oxide layer $d_0$, which is formed during the complete oxidation of a copper film of thickness $h$. Since two copper atoms are consumed to form one $Cu_2O$ molecule, the increase in the number of oxide molecules $\Delta N_{Cu_2O}$ and the decrease in the number of copper atoms $\Delta N_{Cu}$ are related by the equation $Z\Delta N_{Cu_2O} = -\Delta N_{Cu}$, where $Z = 2$ is the stoichiometric coefficient. From here we find the relationship between the increase in the thickness of the oxide layer $\Delta d$ and the decrease in the thickness of the copper film $\Delta h$: $\Delta d = -\gamma \Delta h$, whose integration yields the desired expression $d_0$ through $h$: $d_0 = \gamma h$, where $\gamma = [\rho_{Cu} M_{Cu_2O}]/[Z\rho_{Cu_2O} M_{Cu}]$. In this formula, $\rho_i$ and $M_i$ are the densities and molar masses of copper and its oxide, $i = \{Cu, Cu_2O\}$. Substituting all the necessary parameters into the expression for $\gamma$, we obtain that $\gamma \approx 1{,}64$. (Evaluating the $\gamma$, we used $\rho_{Cu_2O} = 6.11$ g/cm³ [33]).

Since the oxide growth process obeys parabolic law (4), the characteristic time of complete degradation of a copper film of thickness $h$ is determined by the formula $\tau_0 = (d_0^2 - d_i^2)/k_p \approx (\gamma h)^2/k_p$. For example, for a film with a thickness of 100 Å, the characteristic time $\tau_0$ is about 45 days.

In conclusion of this section, it should be noted that the above-discussed "oxygen" mechanism of oxide film growth on copper works not only in the case of structurally imperfect Cu films. The growth of oxide on thick, crystalline perfect copper layers at high temperatures is also due predominantly to the diffusion of oxygen rather than copper ions, as is often assumed in the literature [30, 34-37].

Thus, in work [30], based on detailed thermogravimetric measurements, it is shown that the growth of the oxide film thickness on crystalline perfect copper layers follows the parabolic law $d(t) = \sqrt{k_p' t/\rho_{Cu_2O}^2}$ in the temperature range of 800 – 1050 °C. The constant $k_p'$ in the parabolic law characterizes the rate of growth of the mass of the $Cu_2O$ film and changes with temperature according to the Arrhenius law. The authors of Ref. [30] came to the conclusion that such growth of the oxide film is caused by the diffusion of copper ions towards the outer surface of the Cu₂O film, followed by the interaction of copper with oxygen. It is easy to show that this conclusion is erroneous.

Indeed, the Arrhenius constant for $k'_p$ measured in [30] is $E_A = 170$ kJ/mol, and therefore $T_A = E_A/R = 20808$ K, where $R$ is the universal gas constant. It is clearly seen that $T_A$, within the measurement accuracy of work [23], coincides with the above-mentioned Arrhenius temperature $T_{a0}$, characterizing the diffusion of oxygen atoms as interstitial impurities in Cu₂O. At the same time, it is known that the Arrhenius constant of copper diffusion in Cu₂O is 119 kJ/mol [24], which differs significantly from the above experimental value $E_A$. Therefore, based on the data provided, it can be stated with confidence that in the temperature range of 800–1050 °C, the growth of the oxide film in accordance with the parabolic law is due mainly to the diffusion of oxygen atoms, and not copper ions.

There are a couple more arguments in favor of the "oxygen" mechanism of Cu₂O growth on crystalline perfect copper layers at high temperatures.

Firstly, it is well known that the atomic radius $r_a$ of a neutral oxygen atom is 0.66 Å, while the $r_a$ of a copper cation is almost one and a half times larger than this value and is equal to 0.98 Å [38]. In addition, the mass of a copper cation is four times greater than the mass of an oxygen atom. For this reason, the diffusion coefficient of atomic oxygen in the Cu₂O lattice significantly exceeds that of copper. This is why the main contribution to $k'_p$ is due to atomic oxygen, which is formed from atmospheric molecular oxygen on the outer surface of the oxide layer and then diffuses as interstitial impurity through the Cu₂O layer towards the copper substrate.

Secondly, using the relationship $k_p = 4rD$ derived above and data on the temperature dependence of the diffusion coefficient $D$ of atomic oxygen from work [23], it is possible to obtain the dependence of the thermogravimetric growth constant $k'_p$ on temperature, which practically coincides with the experimental dependence found in work [30]. The latter has the form of the Arrhenius formula with a characteristic energy $E_A = 170$ kJ/mol and a pre-exponential factor $k_p'^0 = 1.08$ (g/cm²)²/s in the temperature range of 800 – 1050 °C.

Indeed, $k_p' = k_p \rho_{Cu_2O}^2 = 4r\rho_{Cu_2O}^2 D_{00} \exp(-T_{a0}/T)$. We have already noted earlier that $E_A/R \approx T_{a0}$. Taking into account that in the high temperature region $r \lesssim 1$, the pre-exponential factor $4r\rho_{Cu_2O}^2 D_{00} \approx 1.6\, r$ (g/cm²)²/s can be considered to coincide with the above experimental value $k_p'^0$.

### *3.3. Evolution of conductivity of Cu films grown on a Ge sublayer*

In the previous section it was shown that the interaction of copper with atomic oxygen leads to a rapid monotonic degradation of the conductivity of Cu films, described by a parabolic law. Copper films grown on a germanium sublayer demonstrate a completely different behavior of conductivity variation with time.

Firstly, as was found in section 3.1., they degrade much more slowly than simple copper films of the same thickness. Secondly, the evolution of their conductivity is non-monotonic: there are parts on the time dependences of the conductivity of Cu/Ge films in which it increases over time. Later in this section it will be shown that these films owe their unusual properties to germanium.

It is necessary to consider the growth process of Cu/Ge films at the microscopic level in order to understand why these films are so resistant to the influence of atmospheric oxygen. The peculiarity of this process is that a small amount of Ge "floats up" [19] and ends up on the surface of the forming Cu film during the formation of a continuous copper layer on the surface of the germanium sublayer.

Chaotically moving along the film surface, germanium clusters, due to polarization in the electric field, accumulate in the vicinity of various types of inhomogeneities: growth steps, dislocations coming to the surface, that is, in those areas on the surface of the copper film where the electric fields are highest. Thus, all areas on the surface of the film with favorable conditions for dissociative adsorption of oxygen are "covered" by a layer of germanium by the time of completion of the Cu film growth. The concept is presented schematically in Fig. 5.

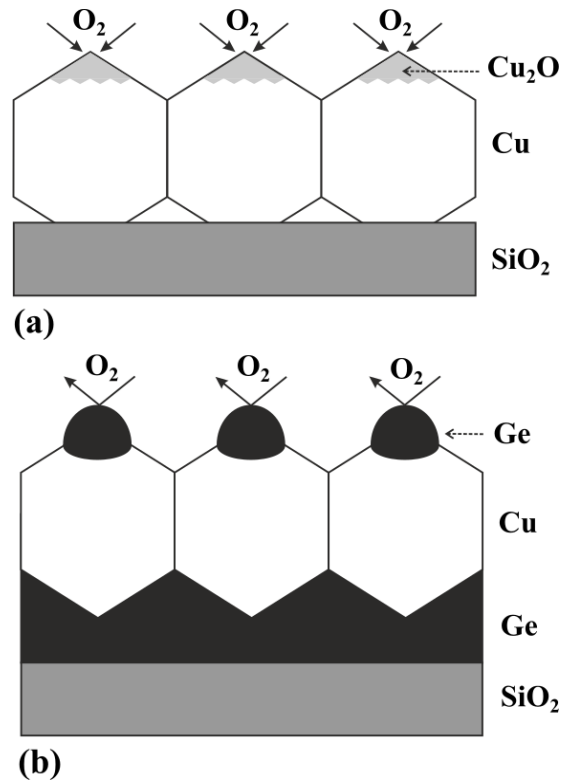

**Fig. 5.** Comparison of the corrosion resistance of Cu and Cu/Ge films. (a) Active dissociative adsorption of $O_2$ and formation of $Cu_2O$ in the vicinity of surface defects of the Cu film, where the electric fields are particularly strong. (b) In the Cu/Ge film, the areas of surface inhomogeneities are covered with Ge, which prevents the adsorption of $O_2$ and thereby blocks the nucleation of $Cu_2O$ islands.

This layer prevents oxygen molecules from getting as close as possible to the inhomogeneities on the metal surface, where the electric fields are particularly strong and the probability of oxygen dissociation is highest. Thus, germanium has a passivating effect on the copper surface, blocking the process of its corrosion at the earliest stage – at the stage of formation of $Cu_2O$ islands. That is why copper films grown on a germanium sublayer are so resistant to the influence of atmospheric oxygen and degrade over time noticeably more slowly than conventional Cu thin films.

The passivating effect of germanium allows one to observe the recrystallization effect in Cu/Ge films during storage, which is usually masked in thin Cu films by their rapid corrosion.

As noted in Section 3.2, thin metal films deposited on substrates at room temperature consist of randomly oriented small crystallites immediately after condensation. However, more and more crystallites become oriented in one direction over time [39]. The dependence of the fraction of such reoriented crystallites $f_o$ on time usually obeys the Avrami equation $f_o = 1 - exp(-\mathcal{K}_a t)$ [40]. A decrease in the degree of crystallite misorientation during film recrystallization leads to an improvement in its conductivity over time. This is precisely what is observed experimentally in the initial part of the $R^\square(t)$ dependences of Cu/Ge films, shown in detail in the inset of Fig. 3.

Since $f_o$ obeys the Avrami equation, the film conductance $G$ changes with time in a similar manner: $G(t) \propto 1 - exp(-\mathcal{K}_a t)$. If we denote the conductance value at $t = 0$ by $G_0$, and its asymptotic value by $G_\infty$, then $G(t) = G_0 + (G_\infty - G_0)[1 - exp(-\mathcal{K}_a t)]$. Therefore, the ratio $\rho(t)/\rho_0$ is described by the function

$$\rho(t)/\rho_0 = [1 + (G_\infty/G_0 - 1)(1 - exp(-\mathcal{K}_a t))]^{-1}, \qquad (5)$$

depending, as can be seen, on two parameters $g = G_\infty/G_0$ and $\mathcal{K}_a$. Fitting this function to the experimental data shown in the inset of Fig. 3 allows one to obtain reliable values of the Avrami constants $\mathcal{K}_a$ for three Cu/Ge samples differing in the thickness of the copper film.

Table 1. The parameters $g$ and $\mathcal{K}_a$ of Avrami's model, obtained by fitting equation (5) to the experimental points shown in the inset of Fig. 3.

| Sample | $g = G_\infty/G_0$ | $\mathcal{K}_a$, $10^{-7}$ s$^{-1}$ |
|---|---|---|
| Cu (2.1 nm)/Ge (2.1 nm) | 1,49 ± 0,04 | 16 ± 2 |
| Cu (4.8 nm)/Ge (2.0 nm) | 2,9 ± 0,9 | 3 ± 1 |
| Cu (7.3 nm)/Ge (2.0 nm) | 1,09 ± 0,02 | 2,6 ± 0,6 |

The values of $\mathcal{K}_a$ obtained in this way are given in Table 1. It is clearly seen that the values of the Avrami constants decrease with increasing thickness of the copper layer. Since with increasing thickness of Cu films the values of $\mathcal{K}_a$ decrease and the average sizes of crystallites increase [41, 42], it should be concluded that smaller crystallites reorient themselves during the recrystallization process faster than larger ones.

The reasons for this effect, as well as many other phenomena in the Cu/Ge system, including its degradation, are in the features of Ge diffusion along the intercrystalline boundary of copper films.

The diffusion of Ge inside thick perfect copper layers was comprehensively investigated by Lohmann et al. [43]. The rate of diffusion propagation of Ge along the intercrystalline boundaries of copper was found to be quite high. If we extrapolate the data from Ref. [43] to room temperature, then the values obtained for the coefficient of diffusion of Ge along the grain boundaries of Cu are in the range from $1.2 \cdot 10^{-17}$ to $2 \cdot 10^{-16}$ cm$^2$/s, which is comparable to the coefficient of diffusion of atomic oxygen along the intercrystallite boundaries of $Cu_2O$ (see section 3.2.).

On the contrary, at room temperature, the diffusion coefficient of Ge in the bulk of Cu grains is many (seventeen! [43]) orders of magnitude less than the coefficient of diffusion of germanium along the grain boundaries. For this reason, the diffusion of Ge along the intercrystalline boundary is essentially one-dimensional. The magnitude and direction of the diffusion flow are determined by the concentration gradient of Ge along the boundary. Therefore, inside the copper film condensed on the germanium sublayer, there is a flow of Ge atoms moving along the intercrystallite boundaries in the direction of the outer surface of Cu film, where the Ge concentration is not as significant as in the sublayer.

Germanium passes with difficulty from the intercrystalline space into the bulk of grains owing to the huge difference in the magnitude of its diffusion coefficients inside the Cu grain and along the grain boundary. Therefore, during the diffusion process, a thermodynamically nonequilibrium concentration of Ge is established quite quickly at the boundaries of Cu grains. This leads to the appearance of pressure acting on the grains, the nature of which is similar to the reason for the appearance of osmotic pressure in solutions [44]. Estimates made in the work by

Bokshtein et al. [44] show that the pressure experienced by copper grains can reach a hundred of atmospheres, depending on the degree of non-equilibrium of the Ge concentration.

Due to the high pressure, a large number of dislocations are formed in the areas adjacent to the boundary of the Cu grains, which then spread into the depths of the grains (crystallites). The formation and movement of dislocations, in turn, causes an easier transfer of germanium along dislocation tubes from the intercrystalline space into the bulk of Cu grains. This, firstly, helps to reduce the degree of nonequilibrium of Ge concentration at grain boundaries, and secondly, leads to a decrease in the intracrystallite conductivity of copper [45].

The appearance of pressure caused by the nonequilibrium concentration of Ge in the grain boundary region allows us to understand the features of the recrystallization process and largely determines the mechanism of degradation of Cu/Ge films.

Indeed, the deviation of the Ge concentration in the grain boundary region from the equilibrium value is quite small after not too much time passed since the condensation of the copper film on the germanium sublayer. As long as the nonequilibrium of the Ge concentration is small, the forces acting on the crystallites due to the "osmotic" pressure facilitate their reorientation from metastable to more stable positions. This results in some improvement in the conductivity of Cu/Ge films simply due to the denser packing of copper grains and the increase in the number of contact points between them. These considerations allow us to explain some increase in the conductivity of Cu/Ge films observed in the initial parts of the experimental dependences $R^{\square}(t)$, shown in the inset of Fig. 3.

However, later, as the nonequilibrium of the Ge concentration increases, the forces acting on individual crystallites increase at some point to such an extent that it becomes possible for the crystallites to be pulled apart from each other (Fig. 6). It is clear that this process will lead to a drop in the conductivity of Cu/Ge films due to the breaks in contacts between previously contacting crystallites. This allows us to explain at a qualitative level why, following the improvement of the conductivity of Cu/Ge films, the process of their steady degradation begins (see Fig. 3 and its inset).

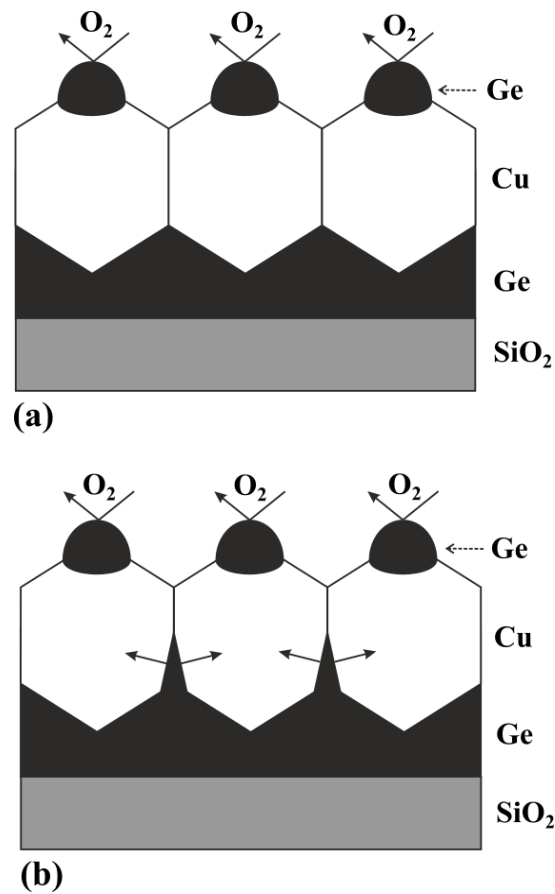

**Fig. 6.** Evolution of the Cu/Ge film structure. (a) The structure immediately after obtaining the film. (b) Destruction of contacts between adjacent Cu crystallites, caused by the action of "osmotic" pressure proposed in the work by Bokshtein et al. [44].

There are two possible main scenarios for the degradation of Cu/Ge films, depending on the ratio between the thicknesses of the germanium and copper layers.

The first of them is realized when the amount of germanium is so great that the processes of its passing into the grains following dislocations are not able to reduce the degree of non-equilibrium of its concentration at the boundary. In this case, the forces acting on the crystallites will steadily pull them apart, destroying the contacts between them. The rupture of an increasing number of contacts between crystallites will at some point lead to a complete loss of conductivity by the Cu/Ge film. Such a scenario of rapid degradation is realized, for example, in a Cu/Ge sample with copper and germanium layer thicknesses equal to 21 Å (see Fig. 3).

The second scenario of Cu/Ge film degradation is realized when the thickness of the copper film significantly exceeds the thickness of the germanium sublayer. In this case, the process of Ge passing from the intercrystalline region into the volume of grains following dislocations will lead at some point to a decrease in the degree of nonequilibrium of its concentration at the grain boundaries. This will lead to a decrease in the forces tending to disrupt contacts between adjacent crystallites due to a decrease in the pressure experienced by them. At some point, this force will become less than a certain characteristic value, below which the process of separating the grains from each other becomes impossible. From this point on, the decrease in film conductivity will occur only due to the penetration of Ge into the copper grains [45] through dislocation tubes. This will lead to a sharp slowdown in the degradation process of Cu/Ge films and a qualitative change in the type of dependence of their conductivity on time: a kink in the dependence $R^{\square}(t)$ will appear. As can be seen from Fig. 3, such a degradation scenario is observed for a sample with copper and germanium layer thicknesses equal to 73 and 20 Å, respectively.

It is clear that the determination of the conditions under which a sharp slowdown in the degradation process of Cu/Ge films occurs, is of great practical importance. As follows from the above examination of the main degradation scenarios for such films, to obtain Cu/Ge coatings that retain their protective properties for a long time, it is necessary to select a copper layer thickness four to five times greater than the thickness of the germanium sublayer.

It should be noted in conclusion of this section that the above-described "osmotic" mechanism of Cu/Ge film degradation ceases to work when the germanium concentration at the intercrystalline boundary reaches a thermodynamically equilibrium value. However, even in this case, several factors can also be mentioned that can lead to the degradation of Cu/Ge films due to the peculiarities of the copper-germanium system.

The point is that the penetration of germanium into copper grains through dislocation tubes in combination with the saturation of the intercrystalline space with germanium can lead to degradation of the conductivity of the Cu/Ge film in the long term.

Indeed, germanium has a fairly high dielectric constant ($\epsilon_{Ge} \approx 16$) [46], so its penetration into the space between adjacent crystallites reduces the adhesive forces between them. On the other hand, the penetration of Ge into copper crystallites leads to the formation of double electrical layers on the opposite sides of the intercrystallite boundary, and these double layers repel each other [47]. The presence of these two processes can lead to the repulsion of adjacent crystallites and the expansion of the intercrystalline space between them, which will cause degradation of the conductivity of the Cu/Ge film.

## 4. Conclusion

In this article, the influence of a germanium sublayer on the native corrosion of ultra-thin copper films with a thickness of about 10 nm was studied for the first time. To this end, we compared the degradation rates of the microwave reflectivity and electrical resistivity of Cu/Ge films and simple copper films deposited directly onto quartz glass substrates. The study of each type of film led us to discover completely new facts regarding the process of corrosion of copper layers, which has been studied for over a hundred years.

Perhaps the most striking effect revealed is the anomalously slow degradation of Cu/Ge films. We found that they corrode significantly more slowly than simple copper films of similar thickness. This is due to the fact that during the growth of the copper film on the germanium sublayer, a small amount of Ge appears on the surface of the forming Cu film, where it accumulates in areas with favorable conditions for the dissociative adsorption of oxygen. By preventing the formation of atomic oxygen in these areas, germanium has a passivating effect on the surface of copper, blocking the corrosion process at the earliest stage – at the stage of formation of $Cu_2O$ islands.

The long-term retention by Cu/Ge films of their characteristics is a practically important and valuable fact, which allows them to be recommended as a replacement for gold coating in EMI protection devices operating at room temperature. Since Cu/Ge films are much cheaper than gold ones, the goals we set in the article can be considered as achieved.

Despite the fact that Cu/Ge films retain their properties for quite a long time, at some point they also begin to degrade. We found that their degradation process is quite unusual, and its rate strongly depends on the ratio of the thicknesses of the copper and germanium layers. In the article, we showed that such a degradation mode of the protective properties of Cu/Ge films is associated with the high diffusion activity of germanium along the intercrystalline boundaries of the copper films and its extremely difficult penetration into the Cu crystallites. The huge (17 orders of magnitude!) difference in the diffusion coefficients describing these two processes quite quickly leads to the formation of a thermodynamically nonequilibrium concentration of germanium at the grain boundaries and the appearance of pressure acting on the crystallites in accordance with the mechanism proposed by Bokshtein et al. [41].

The dependence of this pressure on the degree of nonequilibrium of the germanium concentration at the grain boundaries allowed us to explain the non-monotonic evolution of the conductivity of Cu/Ge films and the dependence of the rate of their degradation on the ratio of the thicknesses of the copper and germanium layers. Understanding these features allowed us to formulate recommendations for obtaining Cu/Ge coatings that retain their protective properties for a long time.

Thus, our studies revealed the duality of the influence of germanium on the degradation process of Cu/Ge films. It is paradoxical that germanium acts both as an inhibitor of surface corrosion of such films caused by atomic oxygen and as a cause of structural changes in them, leading to the degradation of their conductivity (Fig. 6).

Another remarkable fact we found concerns the oxidation of ultra-thin copper films. As it turned out, at room temperature the growth of the oxide thickness $d$ on such films follows the parabolic law $d^2 \propto t$. This is in contradiction with existing theories of copper oxidation, which predict that under similar conditions $d$ increases with time according to a logarithmic or inverse logarithmic law.

We found that the reason for this discrepancy is in the high structural imperfection of ultra-thin copper films. This leads to the fact that the oxide layer growing on the surface of such films is also highly structurally imperfect and saturated with a large number of pores, forming a developed system of channels through which facilitated transport of atomic oxygen occurs from the air/oxide interface to the surface of copper. In the article, we showed that the model linking the growth of the oxide layer with the diffusion of atomic oxygen along the boundaries of $Cu_2O$ grains leads to a parabolic law, reasonable values of the oxide layer thickness and the time of complete oxidation of the film $\tau_0$.

Since metal layers in EMI protection devices under industrial conditions are deposited in a technical vacuum and are therefore usually fine-crystalline and structurally imperfect, the model proposed by us also has practical significance, since it can be used to evaluate a number of parameters that are necessary in practice. For example, using $\tau_0$ it is possible to evaluate the service life of devices that include ultra-thin copper films.

**Acknowledgements**

The authors are grateful to I. Khorin, who informed us about work [19]. We thank P. Glazunov for criticism of the initial version of the article, Yu. Pinaev for performing microwave measurements, and R. Denisov for computer rendering of Figs. 5 and 6. This work was supported by the Russian Ministry of Science and Higher Education through the state assignment of the Kotelnikov Institute of Radio